\documentclass[a4paper,11pt]{article}

\usepackage{jinstpub}
\usepackage{threeparttable}

\title{Pre-launch characterization of the spectrometer of Hard X-ray Imager (HXI) onboard the ASO-S mission}

\author[a,b]{W.~Liu,}
\author[a,b,1]{Z.~Zhang\note{Corresponding author.},}
\author[a,b]{J.~Wu,}
\author[a]{T.~Ma,}
\author[a]{Y.~Zhang,}
\author[a]{Y.M.~Hu,}
\author[a]{Y.Q.~Zhang,}
\author[a]{Y.Y.~Huang,}
\author[a,b]{H.X.~Wang,}
\author[a,b]{X.K.~Jiang}
\author[a]{and D.Y.~Chen}

\affiliation[a]{Key Laboratory of Dark Matter and Space Astronomy, Purple Mountain Observatory,\\
Chinese Academy of Sciences,\\
No. 10 Yuanhua Road, Nanjing 210023, China}
\affiliation[b]{School of Astronomy and Space Science, University of Science and Technology of China,\\
No. 96 Jinzhai Road, Hefei 230026, China}

\emailAdd{zhangzhe@pmo.ac.cn}

\abstract{As one of the three payloads of the Advanced Space-based Solar Observatory (ASO-S), the pre-launch characterization of HXI includes the characterization of the collimator and the spectrometer. This article focuses on the pre-launch characterization of HXI's spectrometer, including detection efficiency, energy resolution, and energy to ADC channel (E-C) relationship. The detection efficiency characterization necessitates a sufficient number of calibrated energy points in the HXI observation energy range, with a refined measurement around the absorption edge of the lanthanum bromide crystal, and is calibrated with an X-ray beam. The characterization results are consistent with the simulations and show that the detection efficiency difference between detector modules is controlled within $\pm3\%$. Radioactive sources $^{133}\textrm{Ba}$ and $^{137}\textrm{Cs}$ are used to calibrate the energy resolution and E-C relationship. The characterization results show that all detector modules' energy resolutions are better than 26\% at 32 keV, meeting the design specification. And their E-C relationships vary regularly with voltage. All of the characterization results indicate that the flight model of the HXI spectrometer meets the performance requirements and is capable of performing on-orbit observation activities.}


\begin{document}
\maketitle
\flushbottom

%
\section{Introduction}           
\label{sect:intro}

The Advanced Space-based Solar Observatory (ASO-S) is a comprehensive solar mission planned for launch in October 2022 by China. Its scientific goals include simultaneously observing the solar magnetic field, solar flares, and coronal mass ejections (CMEs) and exploring the relationships between them. The ASO-S is outfitted with three scientific equipment to achieve these objectives, the Full-disk vector MagnetoGraph (FMG) for observing the solar magnetic field, the Lyman-alpha Solar Telescope (LST) for observing CMEs, and the Hard X-ray Imager (HXI) for observing solar flares~\cite{gan2019advanced}. 

HXI aims at measuring the non-thermal process in the formation of flares by imaging with solar hard X-rays~\cite{su2019simulations}. HXI adopts a spatial modulating Fourier-transformation technique that is often used in the same kind of high-energy imaging instruments, like YOHKOH/HXT~\cite{kosugi1991hard}. HXI is composed of three parts, a collimator (HXI-C) for modulating incident X-rays, a spectrometer (HXI-S) for collecting and recording information from these X-rays, and an electronic Control Box (HXI-E) for data processing and power supply of HXI. 

HXI-S is made up of an array of Lanthanum Bromide (LaBr$_3$(Ce)) scintillator detectors. It is critical for the HXI flight model that the performance of HXI-S should be well calibrated and evaluated before launch to ensure its on-orbit observation. As a result, the characterization of HXI-S is an important part of pre-launch test works. Meanwhile, pre-launch characterization will provide actual HXI-S characteristics as well as be employed for on-orbit data processing. HXI-S pre-launch characterizations include energy linearity, energy resolution, detection efficiency, and so on. In addition, since the detector arrays of HXI-S need to work together to monitor the X-ray from the Sun, the uniformity of the detector modules need to be controlled during the manufacturing process. For example, the disuniformity in energy resolution between different detector modules is required to be less than 10\%, and the difference in detection efficiency is required to be less than 5\%. Similar characterizations have been accomplished for the HXMT and CECAM spacecraft, which adopt NaI(Tl) and LaBr$_3$ scintillators as detectors, respectively~\cite{li2019ground,he2021ground}.

This paper will introduce the pre-launch characterization of HXI-S in detail. As a background, HXI-S, especially the detector module, will be described first. Then the contents about detection efficiency, energy resolution, and energy to ADC channel (E-C) relationship will be presented respectively, including essential principles, facility, processes, and results of each characterization. A brief summary of the characterization will be given at last.

\section{HXI-S and detector module}
\label{sect:HXI-S}

\begin{figure} [htbp]
  \centering
  \includegraphics[width=10.0cm, angle=0]{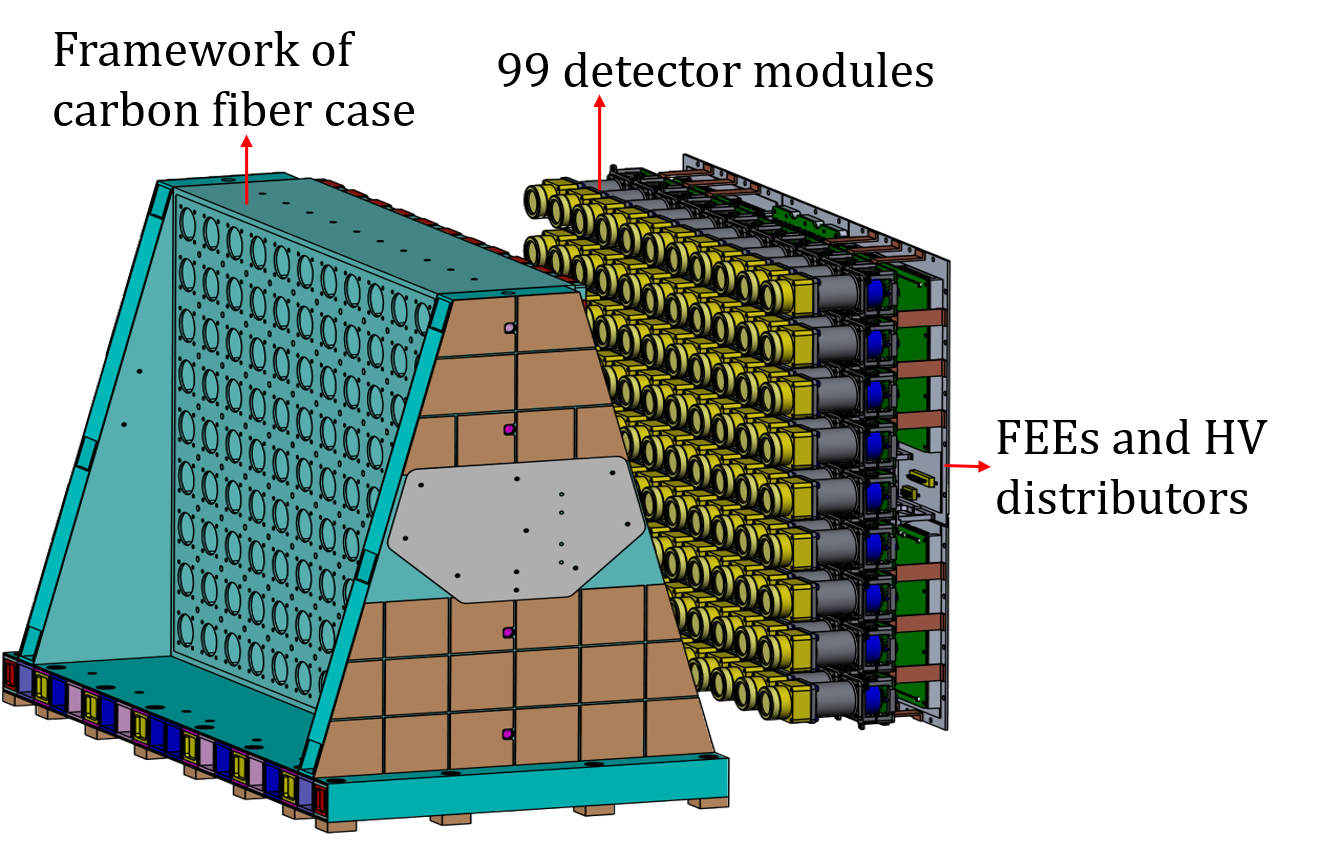}
  \caption{Exploded view of HXI-S.} 
  \label{Fig1}
  \end{figure}

A carbon fiber reinforced plastic (CFRP) framework, 99 detector modules, eight sets of front-end electronics (FEEs), and high-voltage distributors make up HXI-S, as shown in figure~\ref{Fig1}. The framework is in charge of supporting and protecting the detector modules and the external electronics. A LaBr$_3$ scintillator and a photomultiplier tube (PMT) are included in each detector module~\cite{zhang2019hard}. It is used to record the X-rays flux modulated by the sub-collimator in front. Signals from detector modules will be collected and processed briefly by FEEs before being transferred to HXI-E. Table~\ref{tab1} lists the major performance specifications of HXI-S. 

\begin{table}[htbp]
\centering
\caption{\label{tab1} Main performance specifications of HXI-S}
\smallskip
\begin{threeparttable}
\begin{tabular}{lc}
\hline
Energy range & $30\sim200$ keV \\
Energy resolution (FWHM\tnote{1} ) & $\leq 27\% @ 32$ keV \\ 
Time resolution & $\leq 0.5$ s \\
Dead time & 2 $\mu$s \\
\hline
\end{tabular}
\begin{tablenotes}
\footnotesize
    \item[1] FWHM is the full width at half-height of the full-energy peak in the spectrum.
\end{tablenotes}
\end{threeparttable}
\end{table}

\begin{figure} [htbp]
   \centering
   \includegraphics[width=12.0cm, angle=0]{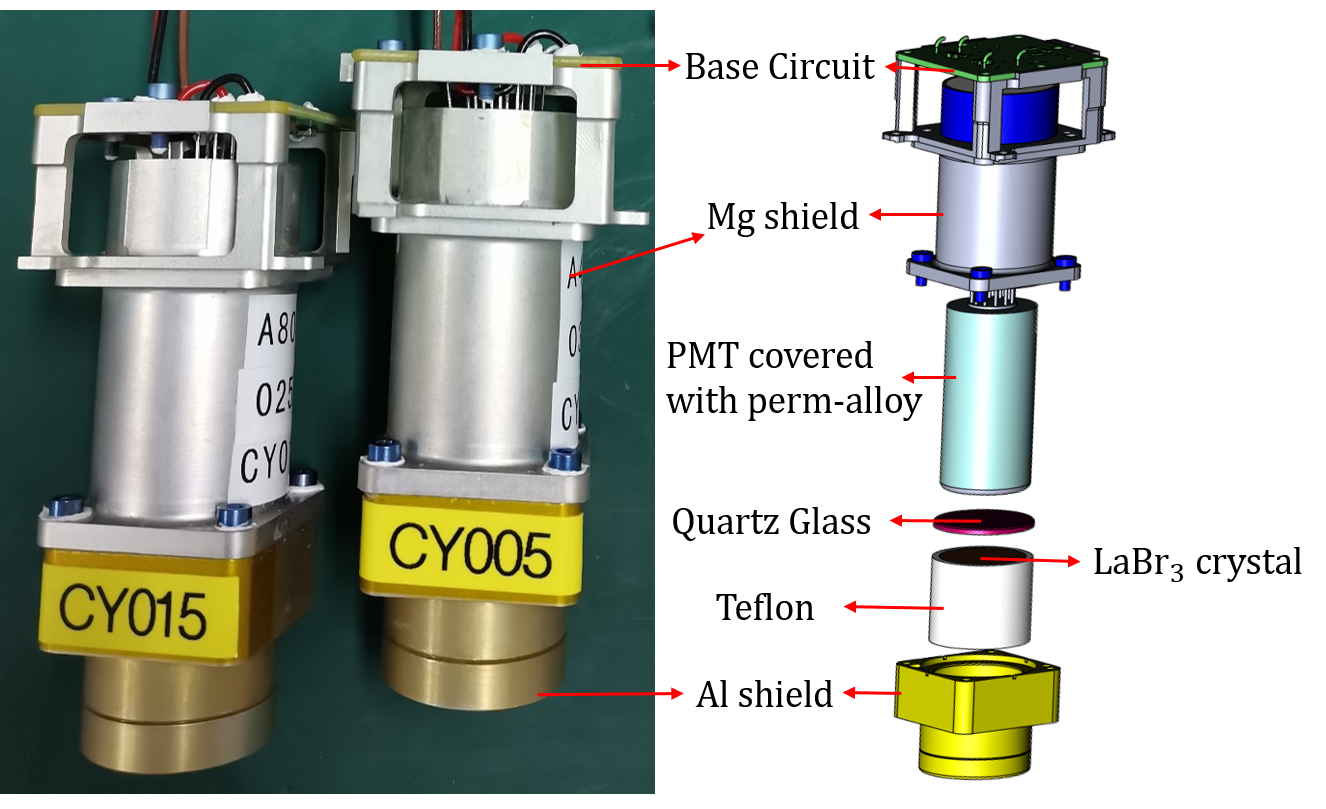}
   \caption{Physical and exploded drawing of LaBr$_3$ detector module.} 
   \label{Fig2}
   \end{figure}

HXI-S detector module is made up primarily of a LaBr$_3$ scintillator, a PMT, and a base circuit, shown in figure~\ref{Fig2}. The LaBr$_3$ crystal provides excellent detection features, such as high light yield, short dead time, outstanding energy resolution, and consistent light output~\cite{moszynski2006temperature}. Due to its excellent properties, LaBr$_3$ scintillator detectors are used by many space science missions~\cite{ma2013gamma, he2021ground}. In the crystal, the energy released following the interaction of photons is transformed into fluorescence, which is subsequently turned into electrons at the photocathode of PMT. The electrons are then amplified in the PMT, and finally, a voltage pulse is generated. The pulse signal is sent to the FEEs via the electronic base circuit. FEEs will amplify and digitize the pulse signal before sending it to HXI-E. More descriptions of electronics can be found in related articles~\cite{wan2022design}.

The LaBr$_3$ crystal is doped with 0.5\% Cerium, its size is $\Phi$25 mm$\times$25 mm, and it is manufactured by the Beijing Glass Research Institute (BGRI). As illustrated in figure~\ref{Fig2}, the crystal is encased in Teflon film and protected by an aluminum shield. Behind the crystal is a 3 mm thick quartz pane that transmits fluorescence. The PMT, R1924A-100, produced by Hamamatsu Ltd, uses a super bi-alkali cathode to ensure good energy resolution. Three layers of perm-alloy foils cover the PMT, reducing the influence on gain variation by the magnetic field~\cite{chen2020environmental,zhang2019hard}. The base circuit for PMT is to maintain linearity and gain stability at a high event rate, and relevant tests can refer to~\cite{zhang2019spectrometer}. Regarding the gain variation of PMT, some explanations are given here. The gain of PMT decreases as the temperature increases, however the thermal control system of the satellite will keep the whole detector of HXI in a stable temperature range at 22$\pm$5$^{\circ}$C, so as to not induce any noticeable gain drift. On the other hand, the reduction of detector gain due to the aging of PMTs after a long of time observation in orbit can be compensated by adjusting the voltage higher.

\section{Characterization of detection efficiency}

The detector characterization procedure establishes an accurate relationship between the input signal and the output signal, ensuring that the input signal can be reliably recovered from the output signal. Monoenergetic X-ray with good collimation is often required for such detector characterization. There are two common methods to obtain monoenergetic X-rays: one is characteristic X-rays from radioactive isotope decay, and the other is monoenergized continuous X-rays generated by an X-ray machine or synchrotron radiation. The former method is simpler, but the number of X-ray energy points is limited. Therefore, the latter method has been used more and more in detector characterization.

Without taking into consideration the influence of electronic signal pile-up, which occurs only when the energetic particle flux is strong enough, the detection efficiency of each detector module is an intrinsic characteristic, hence the match between the FEEs and the detector module is ignored. Therefore, before HXI-S assembly, we can calibrate the detection efficiency. The hard X-ray Calibration Facility (HXCF)~\cite{zhou2014introduction} at the National Institute of Metrology (NIM) of China has been employed in the pre-launch characterization of detection efficiency of HXI-S, which was also used in the calibration studies  of HXMT and GECAM missions~\cite{li2019ground,he2021ground}. Characterization of energy linearity and energy resolution is performed after HXI-S has been assembled. Because the experimental conditions of HXCF do not allow for the testing of the entire HXI-S, and the characterization of these two items does not require precision energy points, we directly use radioactive sources to calibrate the energy linearity and energy resolution, which will be discussed further in the section~\ref{sect:Cal}.


\subsection{Facility for detection efficiency characterization}

The HXCF, which employs the Bragg diffraction principle~\cite{cowley1995diffraction} to select the desired single-energy X-rays from the continuum, has better than 0.1\% monochrome performance that is enough to meet the requirements as the source of HXI-S characterization.

\begin{figure} [htbp]
   \centering
   \includegraphics[width=15cm, angle=0]{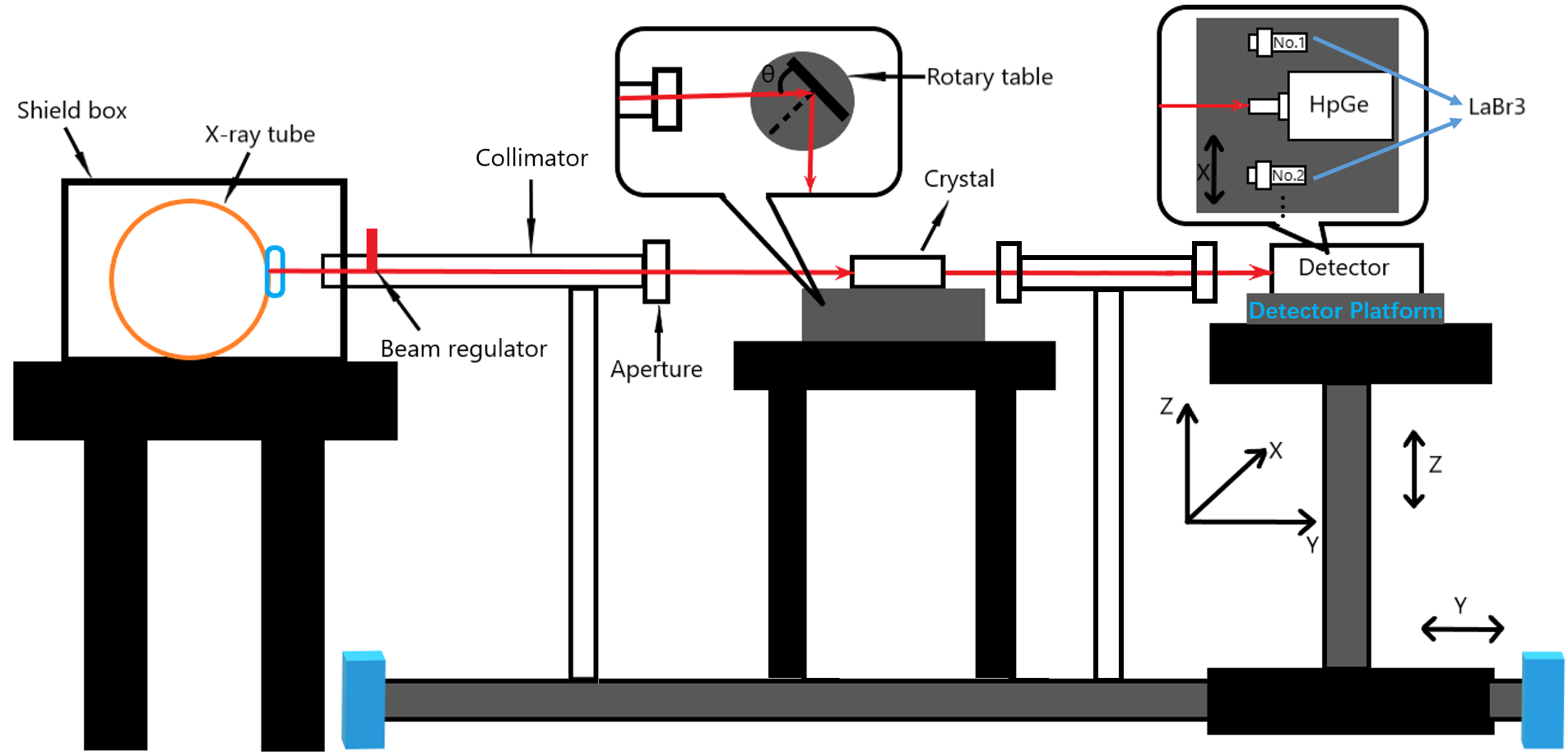}
   \caption{Schematic diagram of the HXCF for the characterization of the detection efficiency of HXI-S.} 
   \label{Fig5}
   \end{figure}

Figure~\ref{Fig5} depicts the HXCF schematic diagram for our characterization experiment. The X-rays released by the X-ray tube are restricted to a tiny beam by the collimation mechanism, including beam regulator, collimator tube, and aperture. A crystal is mounted on a rotator that can be adjusted carefully to keep a specific angle between the crystal and the incident beam. Then a chosen mono-energetic X-ray beam will be produced by Bragg diffraction while the X-ray photons pass through the crystal. At the end of the facility, A detector platform is used to move the detector modules mounted on it on the X, Y, and Z axes to align with the beam. And a high purity germanium (HPGe) detector is also mounted on the platform to provide an absolute detection efficiency reference that will be described in the following subsection. 

\subsection{Procedure of detection efficiency characterization}
\label{sect:Pro}

The characterization experiment's original objective was for the measured energy points to cover the HXI observation energy range (30 keV$\sim$200 keV). However, due to experimental constraints, only the 30 keV$\sim$80 keV energy range was addressed. The energy step in the experiment is set at 2 keV between 30 and 38 keV, 0.1 keV between 38 and 39 keV, and 5 keV between 40 and 80 keV. Because there is an absorption edge between 38 and 39 keV and the efficiency varies dramatically here, the energy interval is designed so tiny that the variation can be measured precisely enough.

The full-energy peak detection efficiency calibrated in this experiment is defined as the ratio of the number of photons with the full energy of photons deposited in the detector's crystal to the number of photons incident on the detector.

The full-energy peak count (FEPC) of each LaBr$_3$ detector module and the HPGe standard detector were measured in the experiment, and the full-energy peak detection efficiency of LaBr$_3$ is as follows:

\begin{equation}
    FEPEff_{LaBr_{3}}=\frac{FEPC_{LaBr_{3}}}{IPC},
\end{equation}

where $FEPEff_{LaBr_3}$ is the full-energy peak detection efficiency of LaBr$_3$, $FEPC_{LaBr_3}$ is the FEPC of LaBr$_3$, $IPC$ is the incident photon count. And the relationship between incident photon count and HPGe is:

\begin{equation}
    IPC=\frac{FEPC_{HPGe}}{FEPEff_{HPGe}}. 
\end{equation}

$FEPC_{HPGe}$ is the FEPC of HPGe, $FEPEff_{HPGe}$ is the full-energy peak detection efficiency of HPGe. Therefore, combining the two formulae above, the full-energy peak detection efficiency of LaBr$_3$ ($FEPEff_{LaBr_3}$) is:

\begin{equation}
    FEPEff_{LaBr_3}=\frac{FEPC_{LaBr_3}}{FEPC_{HPGe}}\times FEPEff_{HPGe}.
\end{equation}

Since $FEPC_{LaBr_3}$ and $FEPC_{HPGe}$ are the experimentally measured quantities, determining $FEPEff_{HPGe}$ can lead to determining $FEPEff_{LaBr_3}$. HPGe is a standard detector that has been calibrated~\cite{haoran2016lege}. The calibration of HPGe, on the other hand, is done with a radioactive source, which has a restricted number of energy points and so does not meet the requirements of this experiment. As a result, Monte Carlo simulation can be used to compare experimental and theoretical results. The simulation software GEANT4 (GEometry ANd Tracking) \cite{agostinelli2003geant4} is used to construct the HPGe mass model, and then the parameters are modified to make the simulated $FEPEff_{HPGe}$ consistent with the radioactive source measurements. The relative deviation between simulated and experimental $FEPEff_{HPGe}$ are within 1.8\%~\cite{haoran2016lege}. Following that, the simulation can be used to determine the various parameters of the HPGe model. The derived HPGe model can be deemed congruent with the actual one. The determined model can then be used to simulate the needed energy point for the characterization experiment. At this point, we may utilize this HPGe model to simulate $FEPEff_{HPGe}$ as the real $FEPEff_{HPGe}$, yielding $FEPEff_{LaBr_3}$ as determined by the experiment: 

\begin{equation}
    ExpFEPEff_{LaBr_3}=\frac{FEPC_{LaBr_3}}{FEPC_{HPGe}}\times SimFEPEff_{HPGe},
    \label{equ7}
\end{equation}

here $ExpFEPEff_{LaBr_3}$ is the $FEPEff_{LaBr_3}$ measured experimentally, $SimFEPEff_{HPGe}$ is the simulated $FEPEff_{HPGe}$.

Following the measurement of $FEPEff_{LaBr_3}$ for each LaBr$_3$ detector module in the experiment, we must simulate $FEPEff_{LaBr_3}$. By comparing the simulation parameters of each detector module to the experimental results, the simulation parameters that best match the actual situation are determined. We obtain the mass model of each detector module in this manner, which we can then utilize for more extensive simulation to obtain the response matrix of each detector module. The response matrix is a two-dimensional correspondence of the detector's response energy spectrum (i.e., output energy spectrum) combined at various input energies, which describes the probability of true spectrum to be reconstructed from response energy spectrum.

\subsection{Experimental and simulation results of detection efficiency}

\begin{figure} [htbp]
   \centering
  \includegraphics[width=15cm, angle=0]{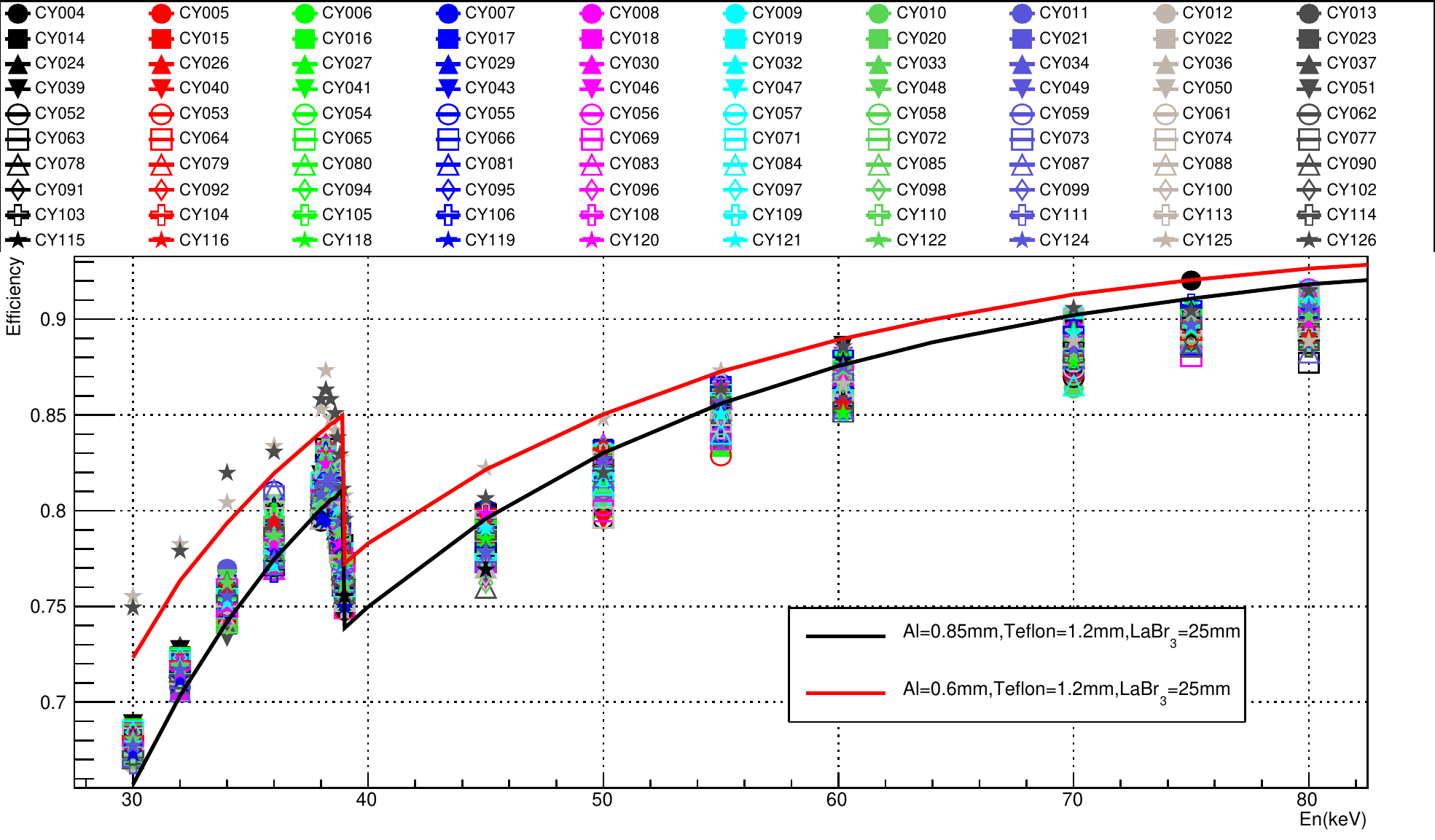}
   \caption{Comparison between experimental and simulated results of full-energy peak detection efficiency. The markers of various colors and forms represent the experimental results of various detector modules, while the solid black and red lines represent the simulated results of various LaBr$_3$ detector model parameters.} 
   \label{Fig8}
   \end{figure}

As shown in figure~\ref{Fig8}, we obtained the final characterization results of full-energy peak detection efficiency by combining the findings of simulation and experiment and applying formula~\ref{equ7}. The simulation model's parameters are the average values supplied in the detector modules' engineering production. We measured 100 detector modules in this experiment, including a backup module. The aluminum shell in front of the detector modules CY125 and CY126 has a design of 0.6 mm thickness, whereas the others have a value of 0.85 mm. All detector modules have the same thickness of Teflon and LaBr$_3$ crystal. The experimental results of detector modules CY125 and CY126 are more compatible with the simulation results of 0.6 mm aluminum shell, while others are more consistent with the simulation results of 0.85 mm aluminum shell, as shown in figure~\ref{Fig8}. Furthermore, the absorption edge from 38 to 39 keV is clearly seen in figure~\ref{Fig8}. The experimental data agree well with the simulated results. 

Although the full-energy peak detection efficiency of energy greater than 80 keV has not been determined due to experimental constraints, the measured full-energy peak detection efficiency is in line with the theoretical value below 80 keV. And, in the energy range of 30 to 200 keV, the interaction between the detector and the X-ray is the same, which means that the performance of each detector module does not vary sharply as the energy increases. As a result, we can simply conclude that the detection efficiency of the full-energy peak at energies more than 80 keV is also commensurate with the theoretical value.

In order for HXI to finally accomplish accurate X-ray imaging, each detector module (except the two thinner ones) should ideally have the same detection efficiency. However, there are usually small variances between detector modules in reality, therefore one of the goals of the characterization experiment is to assess the difference in detection efficiency between detector modules. Then, based on the variations between them, we can adjust their detection efficiency to be consistent. In figure~\ref{Fig8}, the difference in detection efficiency of full-energy peaks between detector modules is not clearly seen. 

\begin{figure} [htbp]
   \centering
  \includegraphics[width=15cm, angle=0]{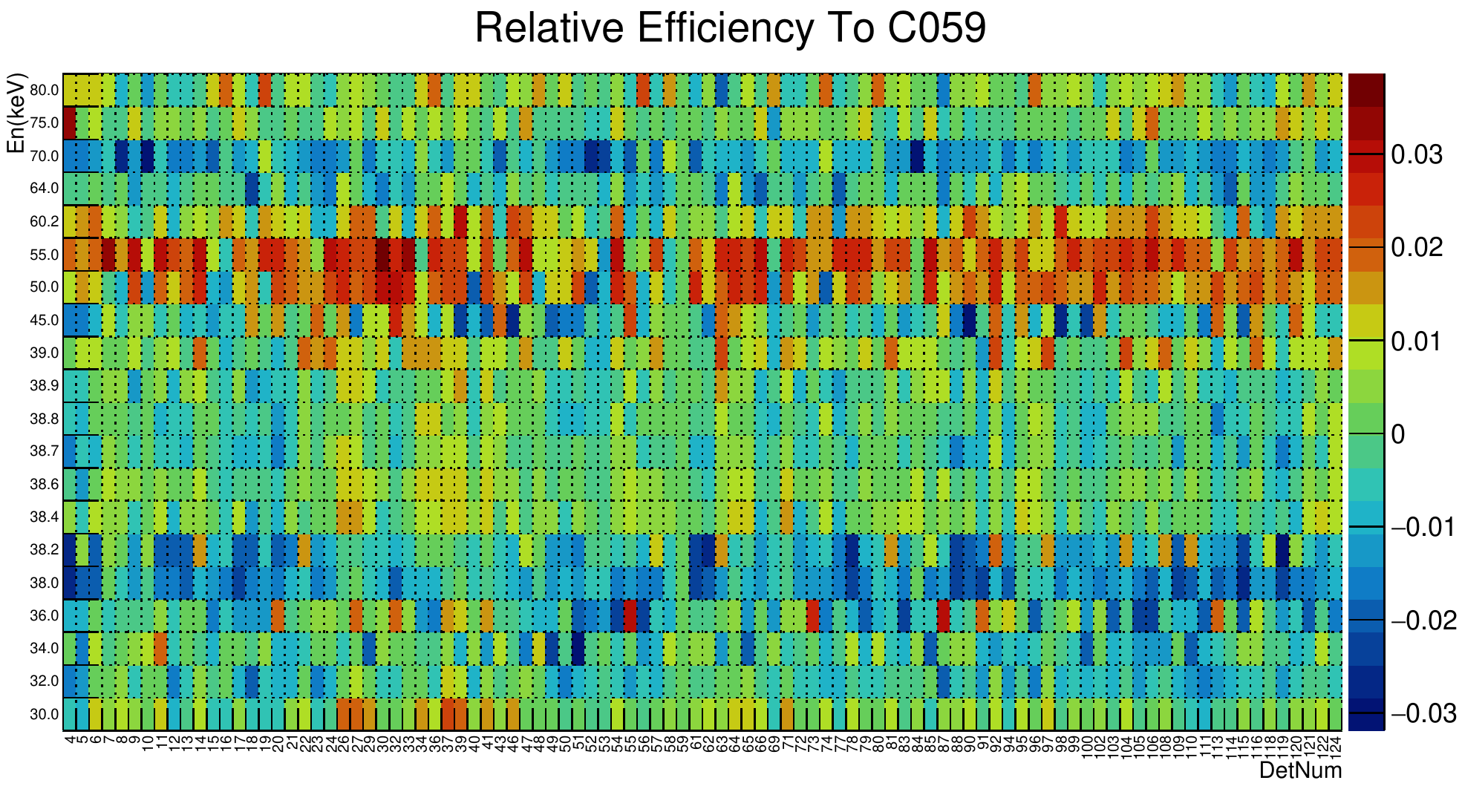}
   \caption{Differences in the detection efficiency of the full-energy peak of different detector modules in comparison to the detector module CY059.} 
   \label{Fig10}
   \end{figure}

Therefore, we choose one of the detector modules (CY059) as a reference to analyze the differences in detector efficiency of different detector modules, as shown in figure~\ref{Fig10}. The detector module CY059 is used to monitor the X-ray flux from the sun. The colorful blocks in the figure reflect the differences in full-energy peak detection efficiency of different detector modules at various energies in comparison to detector module CY059. The difference in full-energy peak detection efficiency between one detector module and the detector module CY059 at energy $E_0$ is

\begin{equation}
    Diff_{OneToCY059}(E_0)=\frac{FEPEff_{One}(E_0)-FEPEff_{CY059}(E_0)}{FEPEff_{CY059}(E_0)},
\end{equation}

here, $FEPEff_{One}(E_0)$ and $FEPEff_{CY059}(E_0)$ are the full-energy peak detection efficiency of one detector module and the detector module CY059 at energy $E_0$, respectively. 

As shown in figure~\ref{Fig10}, the difference in measured full-energy peak detection efficiency between the other detector modules and module CY059 is essentially within $\pm3\%$. It is worth mentioning that the variation of detection efficiency at different energy points exhibits systematic deviations. For example, at 55 keV, the overall efficiency differences are rather large, whereas, at 38 keV, the overall efficiency differences are relatively small. This indicates that the divergence is caused by the test system's stability rather than the detector module itself. However, due to the complexity of the test system, we can only estimate the size of the systematic error here, and correcting the error accurately is an almost impossible task. What's more, the value ($\pm3\%$) is still small enough to meet the detection requirements of HXI.

\section{Characterization of energy resolution and E-C relationship}
\label{sect:Cal}

\subsection{Principle of characterization of energy resolution and E-C relationship}
\label{sect:Pri}

\begin{figure} [htbp]
   \centering
   \includegraphics[width=14cm, angle=0]{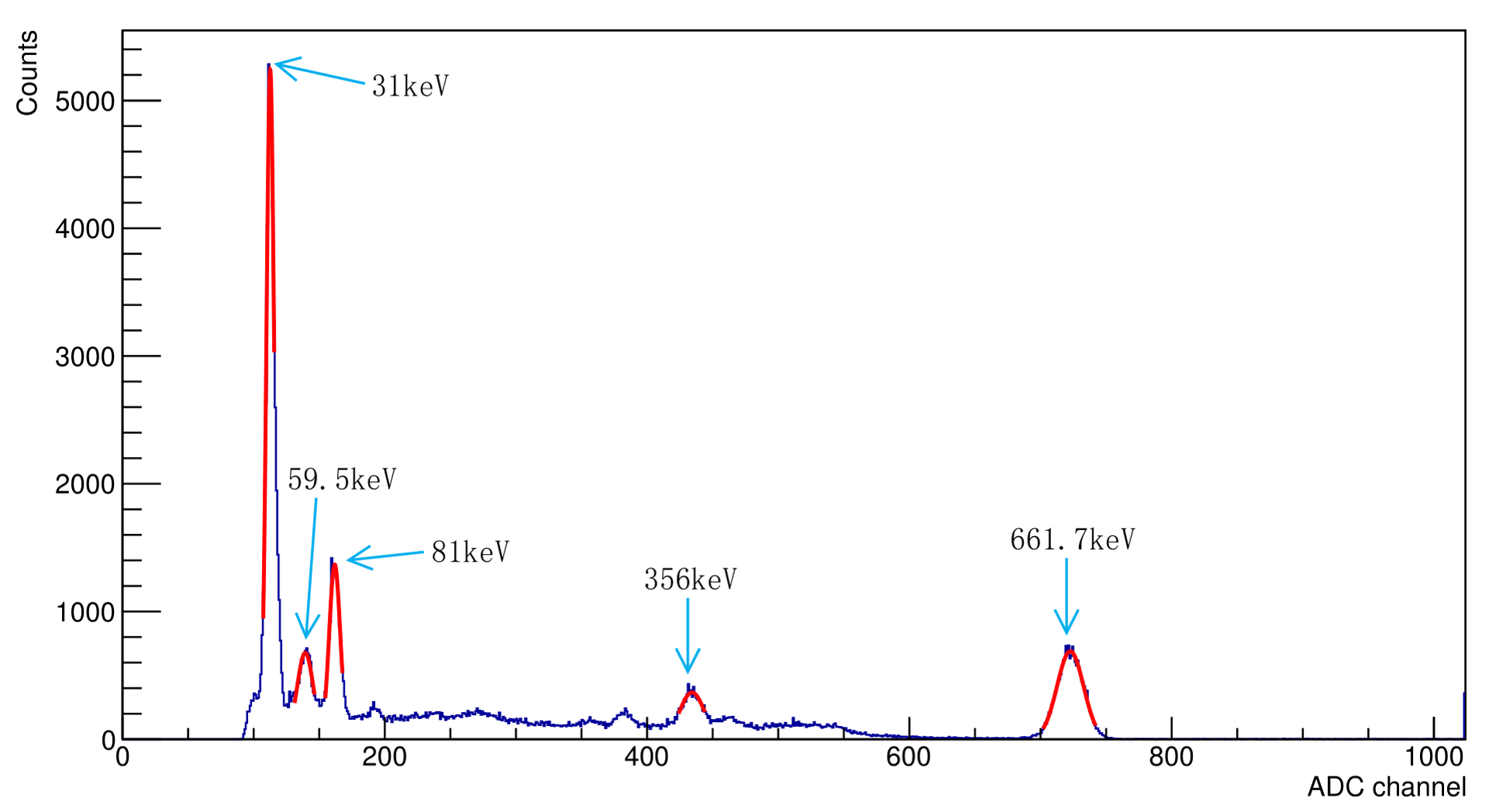}
   \caption{The counting spectrum produced by simultaneously measuring the radioactive sources $^{133}\textrm{Ba}$ and $^{137}\textrm{Cs}$ with the detector module CY063 at 830 V high voltage. The X-axis represents the ADC channel, the Y-axis is counts.} 
   \label{Fig6}
   \end{figure}

The radioactive sources $^{133}\textrm{Ba}$ and $^{137}\textrm{Cs}$ were utilized to calibrate energy resolution and the E-C relationship. Figure~\ref{Fig6} depicts one measured spectrum at normal operating voltage (830 V). This curve contains several full-energy peaks with the following values: 31 keV, 59.5 keV, 81 keV, 356 keV, and 661.7 keV. The full-energy peaks of 31 keV, 81 keV, and 356 keV are mostly caused by $^{133}\textrm{Ba}$, 661.7 keV by $^{137}\textrm{Cs}$, and 59.5 keV by the characterization source $^{241}\textrm{Am}$ embedded in the detector module.

\begin{figure} 
   \centering
   \includegraphics[width=14.5cm, angle=0]{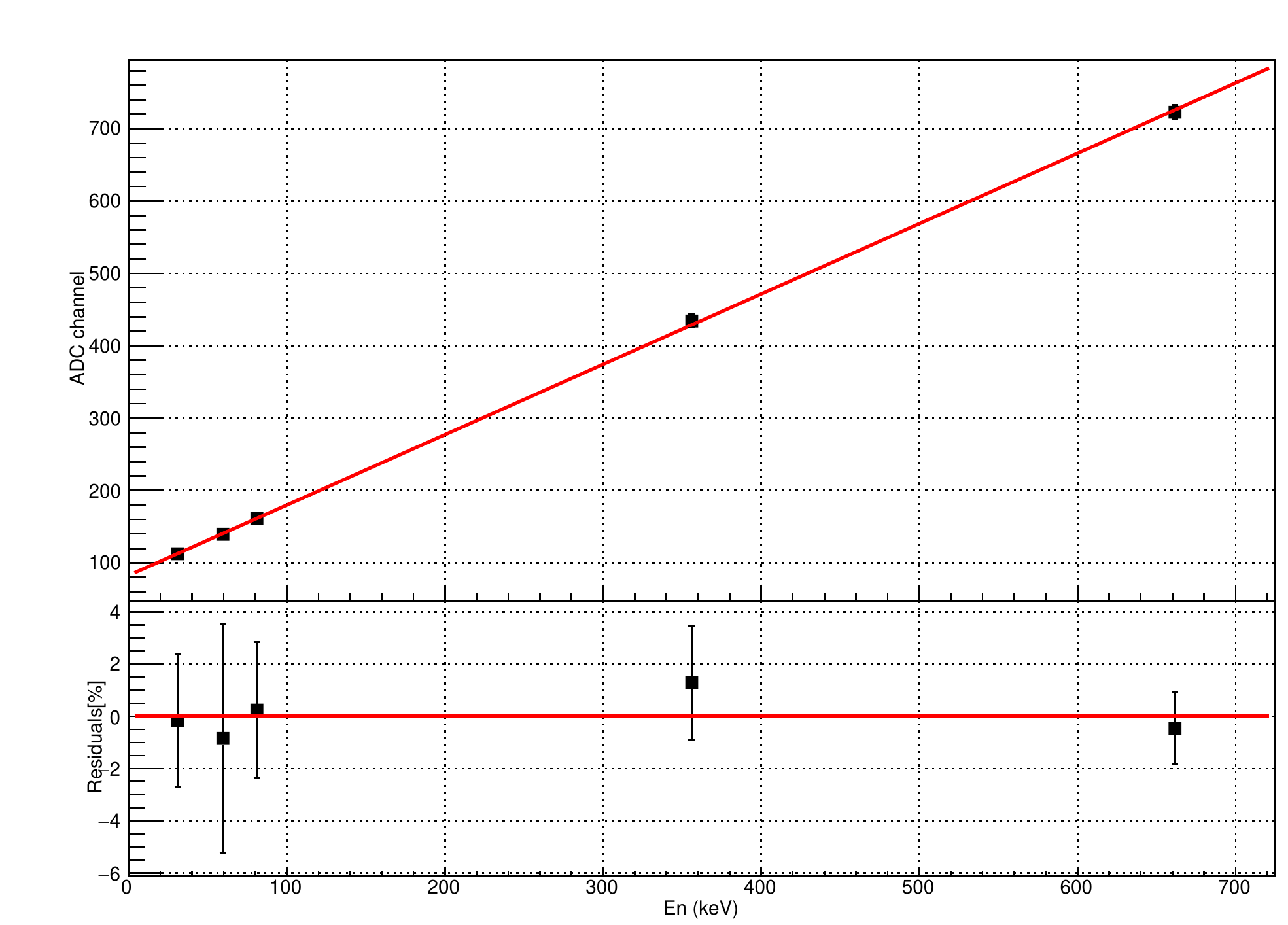}
   \caption{Energy linearity test results of the detector module CY063 at 830 V high voltage.} 
   \label{Fig7}
   \end{figure}

We can determine the number of ADC channels corresponding to their peak positions and standard deviations $\sigma_{ch}$ by fitting the full-energy peaks in the figure with a Gaussian function. The E-C relationship represents the linear relationship between energy and channel and can be expressed as  

\begin{equation}
    ch=A\times E+B,
    \label{equ8}
\end{equation}

where $E$ is energy and $ch$ is the ADC channel as seen in figure~\ref{Fig6}, $A$ and $B$ represent the gradient and intercept, which can be obtained by linear fitting channels and the corresponding energy, as shown in figure~\ref{Fig7}. The gradient $A$ is a quantity related to the electronic gain of the detector module, and its value grows as the provided voltage increases. By varying the amount of the provided voltage, we may investigate the influence of different voltages on the electronic gain of various detector modules.

The energy resolution of detector modules can be expressed as $\eta _E=(2.355\sigma_E)/E$~\cite{mandel2012statistical}, to achieve $\eta _E$, first obtain energy $E$ and its standard deviation $\sigma_E$. Based on formula \ref{equ8}, there is $\sigma_{ch}=A\times\sigma_E$. So the energy resolution can be finally expressed as 
\begin{equation}
    \eta_{E}=\frac{2.355\sigma_{ch}}{A\times E}.
    \label{equ9}
\end{equation}

According to the $\sigma_{ch}$ and $A$ obtained by fitting in figure~\ref{Fig6} and figure~\ref{Fig7} respectively, and the corresponding energy $E$, we will eventually be able to derive the energy resolution $\eta_E$.

\subsection{Result of characterization of Energy resolution and E-C relationship}

\begin{figure} [htbp]
   \centering
  \includegraphics[width=14.5cm, angle=0]{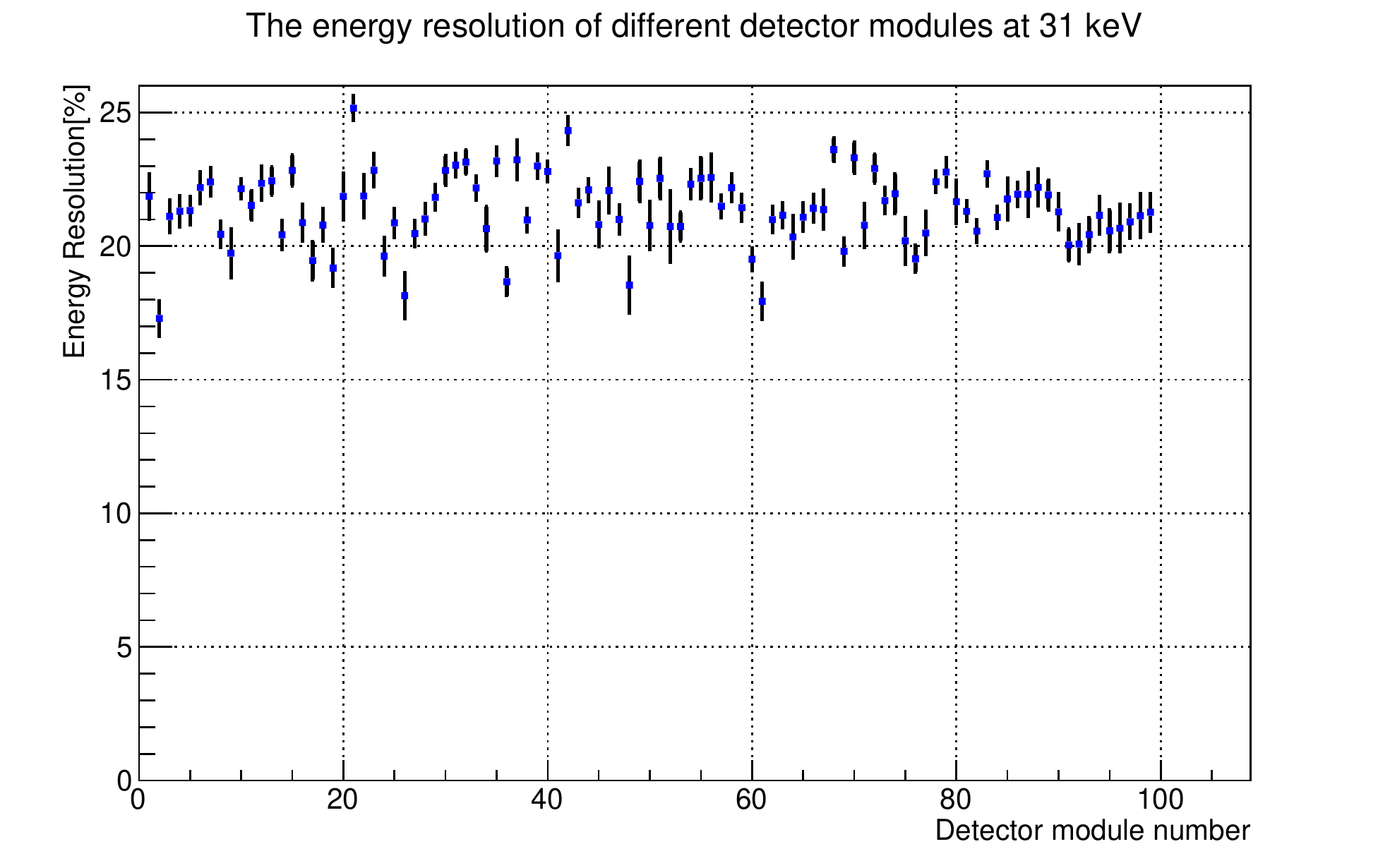}
   \caption{The energy resolution of different detector modules at 31 keV.} 
   \label{Fig11}
   \end{figure}

Table~\ref{tab1} shows the design specification for the energy resolution of HXI-S, which is 27\% at 32 keV. One goal of energy resolution characterization is to guarantee that each detector module's energy resolution meets the design requirements. Using formula~\ref{equ9},  we can calculate the energy resolution at 31 keV for each detector module based on the experimental results. Figure~\ref{Fig11} shows the energy resolution at 31 keV for different numbered detector modules. Since according to figure~\ref{Fig11}, the largest energy resolution of all detector modules is 25.2\%$\pm$0.5\% at 31 keV and it is well known that the percent energy resolution value decreases as the total released energy $E$ increases ($\eta_E\propto1/\sqrt{E}$~\cite{mandel2012statistical}), it can be concluded that the requirement of the resolution being better than 27\% at 32 keV is met. Furthermore, we can see from figure~\ref{Fig11} that the difference in energy resolution between each detector module is less than 8\%, which matches the design criteria.

\begin{figure}
   \centering
  \includegraphics[width=14.5cm, angle=0]{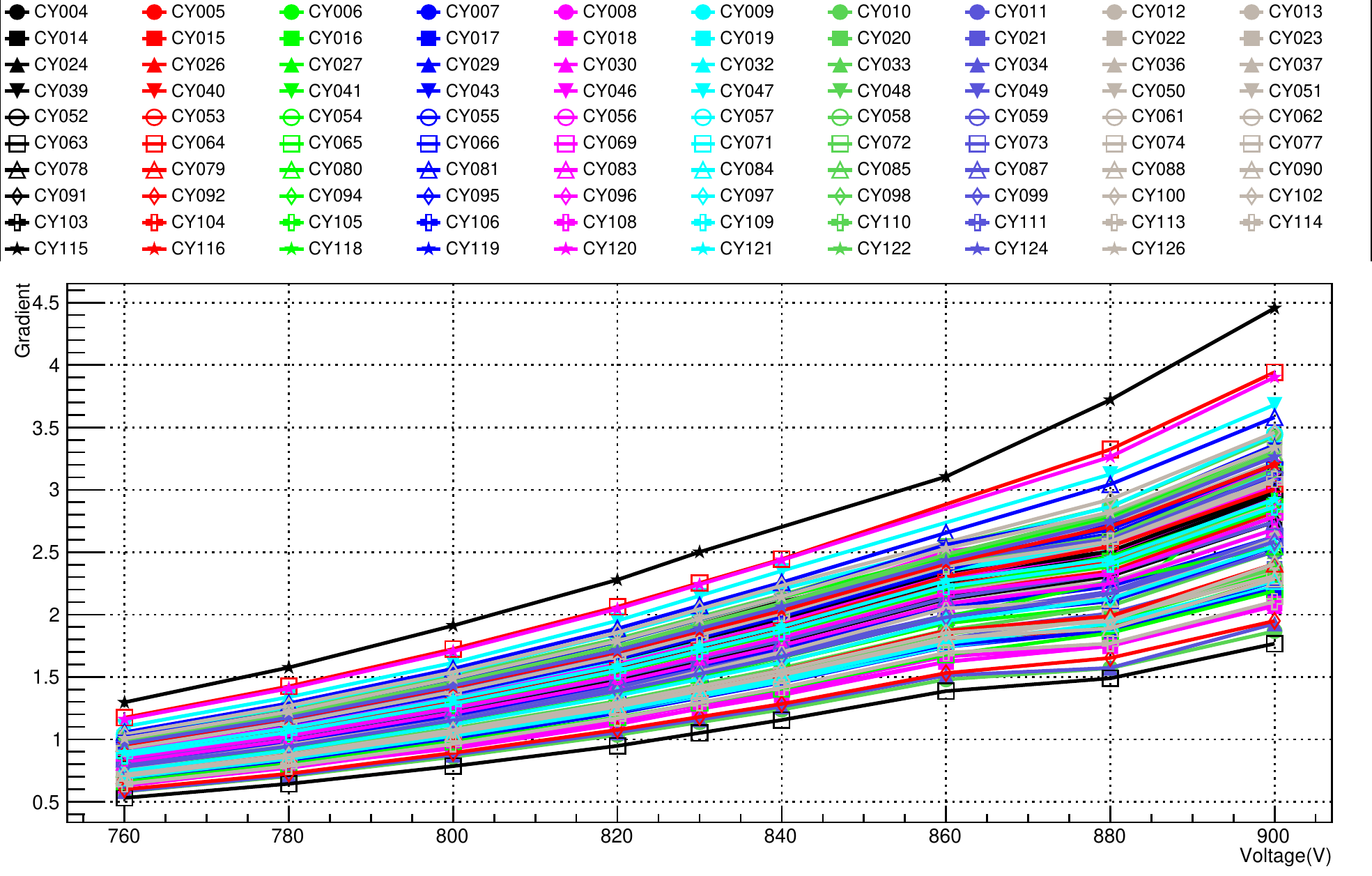}
   \caption{The relationship between electronic gain of different detector modules and supplied voltages.} 
   \label{Fig12}
   \end{figure}

We tested the E-C relationship of all detector modules at various high voltages during the characterization of the E-C relationship. Figure~\ref{Fig12} depicts the influence of varying voltages on the electronic gain of various detector modules. The electrical gain is represented in the figure by the gradient $A$ in Equation~\ref{equ8}. The gradients of all detector modules grow monotonically with voltage and generally follow a similar trend, with the greater the voltage, the faster the electronic gains increase. Furthermore, there is no abrupt change in the electronic gain of all detector modules at varied voltages, showing that the entire instrument can be operated normally at various voltages. Furthermore, the gradient $A$ and intercept $B$ in Equation \ref{equ8} can be determined by linearly fitting the E-C relationship of each detector module at varied voltages. Therefore, the E-C relationship of each detector module at each voltage may be determined, providing the foundation for data support for further satellite observations on orbit.

\section{Summary}

The ASO-S will be launched in the autumn of 2022. The HXI has completed all ground construction and testing. This article discusses HXI-S pre-launch characterization, including full-energy peak detection efficiency, energy resolution, and the E-C relationship.

The detection efficiency characterization experiment was carried out at HXCF of NIM in Beijing. The experiment generated monochromatic X-rays of varying energies using single-crystal Bragg diffraction and the HPGe standard detector as a reference to complete the detector efficiency characterization of all HXI-S detector modules. We acquired the relevant parameters of the appropriate simulation model based on the calibrated experimental data, and we also confirmed that the experimental results are in good agreement with the simulation ones. The experimental data also show the lanthanum absorption edge (38.92 keV) clearly. This is a strong validity confirmation of the test. The difference in full-energy peak detection efficiency of all detector modules at each energy point is controlled within $\pm3\%$.

We use radioactive sources $^{133}\textrm{Ba}$ and $^{137}\textrm{Cs}$ to calibrate HXI-S's energy resolution and E-C relationship. At 32 keV, the energy resolution for all 99 detector modules is better than 26\%, meeting the flight model's requirements. The E-C relationship with voltage variation was also obtained, demonstrating that the entire instrument can operate normally at a variety of voltages. Once the detector gains are altered, the high voltage adjustment on orbit will be greatly aided.

To summarize, we have completed the most critical characterization of HXI-S on the ground. The results of the tests determined the instrument's characteristics, which will be employed for on-orbit data processing.

\acknowledgments

This work is supported by the Strategic Priority Research Program on Space Science, Chinese Academy of Sciences (No. XDA 15320104), National Natural Science Foundation of China (Nos. 11973097, 12173100, and 12022302), and the Youth Innovation Promotion Association CAS (No. 2021317 and Y2021087). We wish to thank the HXI group as well as graduates of NIM, for their valuable work during characterization test. 

\bibliographystyle{JHEP}
\bibliography{Calib}

\end{document}